\documentclass[a4paper,12pt]{article}
\usepackage{epsfig}
\usepackage{graphicx}
\usepackage{dcolumn}
\usepackage{bm}
\usepackage{longtable}
\usepackage{amssymb}
\usepackage{amsfonts}
\usepackage[margin=1.0in]{geometry}
\usepackage{hyperref}
\usepackage{amsmath}


\usepackage{slashed}

\usepackage{hyperref}

\begin{document}
\begin{titlepage}
\begin{flushright}
DISIT-2013\\
CERN-PH-TH/2013-083
\par\end{flushright}
\vskip 1.5cm
\begin{center}
\textbf{\LARGE \bf  A Note on the Field-Theoretical}
\textbf{\vspace{.5cm}}\\
\textbf{\LARGE \bf  Description of Superfluids}

\textbf{\vspace{2cm}}

{\Large L. Andrianopoli$^{~a, c,}$\footnote{laura.andrianopoli@polito.it},
R. D'Auria$^{~a, c,}$\footnote{riccardo.dauria@polito.it},$\ $ \\ \vspace{.5cm}
P.~A.~Grassi$^{~b, d, e,}$\footnote{pgrassi@mfn.unipmn.it},  and
M.~Trigiante$^{~a, c,}$\footnote{mario.trigiante@polito.it}}
\vspace{.5cm}
\begin{center}
 {a) { \it DISAT, Politecnico diTorino, }}\\
 {{
 Corso Duca degli Abruzzi 24,I-10129, Turin, Italy
 }}
\\ \vspace{.2cm}
 {b) { \it DISIT, Universit\`{a} del Piemonte Orientale,
}}\\
{{ \it via T. Michel, 11, Alessandria, 15120, Italy, }}
\\ \vspace{.2cm}
{c) { \it INFN - Sezione di Torino,}}
 \\ \vspace{.2cm}
{d) { \it INFN - Gruppo Collegato di Alessandria - Sezione di Torino,}}
\end{center}
{e) { \it
PH-TH Department, CERN,\\
CH-1211 Geneva 23, Switzerland.}}

\par\end{center}
\vfill{}

\begin{abstract}
{\vspace{.3cm} \noindent } Recently, a Lagrangian description of superfluids attracted some interest from the
fluid/gravity-correspondence viewpoint. In this respect, the work of  Dubovksy {\it et al.} has proposed a new
field theoretical description of fluids, which has several interesting aspects. On another side, we have
provided in \cite{Andrianopoli:2013dya} a supersymmetric extension of the original works. In the analysis of the
Lagrangian structures a new invariant appeared which, although  related to known invariants, provides, in our
opinion, a better parametrisation of the fluid dynamics in order to describe the fluid/superfluid phases.
\end{abstract}
\vfill{}
\vspace{1.5cm}
\end{titlepage}

\vfill
\eject

\newpage
\setcounter{footnote}{0}


Motivated by fluid/gravity correspondence \cite{Bhattacharyya:2008jc,Rangamani:2009xk,Hubeny:2011hd,Bhattacharya:2012zx}
and the recent developments in the field theory description of fluids
by \cite{Brown:1992kc,Dubovsky:2005xd,Dubovsky:2011sj,Nicolis:2011cs,Dubovsky:2011sk}, in \cite{Andrianopoli:2013dya} we have
extended a field-theoretical description of fluid dynamics to a supersymmetric framework. Actually, we considered
the comoving-coordinate description in terms of fields $\phi^I$, adopted first in \cite{Brown:1992kc}\footnote{See also
\cite{Jackiw:2004nm} for a complete review.}, and we
extended it to a supersymmetric framework by introducing a set of fermionic coordinates. In that way, the basic
ingredients of the Lagrangian description are a set of supersymmetric invariants whose bosonic part coincides
with the invariants discussed in the quoted papers. Some of those invariants are also invariant under chemical
shift symmetry, though not all of them. The supersymmetrization requires  to introduce, besides the fermionic
coordinates $\theta^\alpha$, an additional fermionic coordinate $\tau$ partner of  the local coordinate $\psi$
implementing the chemical-shift symmetry, as used in \cite{Dubovsky:2005xd,Dubovsky:2011sj,Nicolis:2011cs}. In
addition, we considered new Poincar\'e
 invariants out of the fields $\partial_\mu \phi^I$ and $\partial_\mu \psi$, $Z^I\equiv \partial_\mu \phi^I \partial^\mu \psi$,
 which  could be useful, together with the other invariants, to describe in a
 more natural way the dynamics of a fluid in some
particular conditions, such as, for example, the superfluid phase transition. A particular combination of the $Z^I$
with the other Poincar\'e invariants is also invariant under   chemical-shift symmetry and provides a natural
understanding of the kinetic term of $\psi$. It actually coincides with $y^2= u^\mu u^\nu \partial_\mu \psi
\partial_\nu \psi$.

In the following, we first give a short review of the Lagrangian formalism for fluid mechanics, with attention
to the chemical shift symmetry and to the derivation of the energy-momentum tensor, where we adhere to the
field-theoretical approach developed in \cite{Dubovsky:2011sj,Nicolis:2011cs}. Then we discuss the properties of
the new Poincar\'e invariant $Z^I$,  presented at the bosonic level in \cite{Andrianopoli:2013dya}, further
extending it to a supersymmetric setting. In the application to the superfluid, it provides a better
parametrization of the superfluid phase separation\footnote{See \cite{Herzog:2011ec,Bhattacharya:2011eea} for 
a review on superfluids in the context of holography and fluid/gravity correspondence}. 
The new invariants would be also useful for the
quantum extension of the theoretical description of fluids along the lines of \cite{Endlich:2010hf}.


The field-theory Lagrangian approach to fluid dynamics was developed in ref.s
\cite{Brown:1992kc,Dubovsky:2005xd,Dubovsky:2011sj,Nicolis:2011cs}. It is based on the use of the comoving
coordinates of the fluid as fundamental fields. We will adopt the same notations as \cite{Dubovsky:2011sj}.


Working, for the sake of generality, in $d+1$ space-time dimensions, one introduces $d $ scalar fields
$\phi^I(x^I,t)$, $I=1,\dots, d $, as Lagrangian comoving coordinates of a fluid element at a point ${x^I}$ and
time $t$, such that a background is described by $\phi^I=x^I$ and requires, in absence of gravitation, the
following symmetries:
\begin{eqnarray}
\delta \phi^I &=& a^I \quad (a^I = \rm const.)\,,\label{symm1}\\
\phi^I& \rightarrow & O^I_J\,\phi^J,\,\,\,\,\quad(O^I_J\,\epsilon \,{\rm SO(d)})\,,\label{symm2}\\
\phi^I&\rightarrow &\xi^I(\phi), \quad\quad \det(\partial \xi^I/\partial \phi^J)=1.\label{symm3}
\end{eqnarray}
The following current respects the symmetries (\ref{symm1}) - (\ref{symm3}):
\begin{equation}\label{correntona}
    J^\mu=\frac{1}{d!}\, \epsilon^{\mu,\nu_1,\dots,\nu_d}
     \epsilon_{I_1,\dots,I_d} \partial_{\nu_1}\phi^{I_1}\dots \partial_{\nu_d}\phi^{I_d}\,,
 \end{equation}
 and enjoys  the important property that its projection along the comoving coordinates does not change:
\begin{equation}\label{enjoy}
   J^\mu\,\partial_{\mu}\phi^{I}=0.
\end{equation}
This is equivalent to saying that the spatial d-form current
$J^{(d)}=-{\star^{{d+1}}} J^{(1)}$, where
\begin{equation}\label{J1}
    J^{(1)}=\frac{1}{d!}\, \epsilon_{\mu \nu_1 \dots \nu_{d}}\,\epsilon_{I_1\dots I_d}\partial^{\nu_1}\,\phi^{I_1}
    \dots \partial^{\nu_d}\,\phi^{I_d}\,dx^\mu =
  (-1)^d\star^{d+1}\,\left(\frac{1}{d!}\,\epsilon_{I_1\dots I_d} d\phi^{I_1}{\wedge} \dots {\wedge} d\phi^{I_d}\right)\,,
\end{equation}
 is closed identically, that is it is an exact form.
Hence it is natural to define the fluid four-velocity as aligned with $J^\mu$:
\begin{equation}\label{align}
    J^\mu=b\,u^\mu \rightarrow b=\sqrt{-J^\mu\,J_\mu}=\sqrt{{\rm det}( {B}^{IJ})}\,,
\end{equation}
where $ {B}^{IJ}\equiv \partial_\mu \phi^I \partial^\mu \phi^J$. From a physical point of view, the property of
$J^\mu$ to be identically closed identifies it with the entropy current of the perfect fluid in absence of
dissipative effects, so that $b=s$,  $s$ being  the entropy density.

If there is a conserved charge (particle number, electric charge etc.),  one introduces a new field
$\psi(x^I,t)$ which is a phase, that is it transforms under ${\rm U}(1)$ as follows
\begin{equation}\label{psitrans}
  \psi\rightarrow  \psi + c, \quad (c= \rm const.).
\end{equation}
Moreover, if the charge flows with the fluid, charge conservation is obeyed
separately by each volume element. This means that the charge conservation is not affected by an arbitrary
comoving position-dependent transformation
\begin{equation}\label{extsym}
   \psi\rightarrow  \psi + f(\phi^I)
\end{equation}
$f $ being an arbitrary function. This extra symmetry requirement on the Lagrangian is dubbed
\emph{chemical-shift symmetry}. Using the entropy current $J^\mu$ one finds that, by virtue of eq. (\ref{enjoy}), the quantity $J^\mu
\partial_\mu\psi$ is invariant under (\ref{extsym}).

From these premises the authors of \cite{Dubovsky:2011sj}  constructed the low energy Lagrangian respecting the
above symmetries. To lowest order in a derivative expansion the Lagrangian will depend on the first derivatives
of the fields through invariants respecting the symmetries (\ref{symm1}) - (\ref{symm3}), (\ref{psitrans}):
\begin{equation}\label{leg}
   \mathcal L = \mathcal L (\partial \phi^I,\partial \psi).
\end{equation}
In principle, there are two such invariants  constructed out of
 $J^\mu$ and $\partial_\mu\psi$, namely $b$ and $J^\mu\partial_\mu\psi$, so that the (Poincar\'e invariant) action functional can be written as follows:
\begin{equation}\label{symmlagr}
   S=\int d^{d+1}\,x F(b,y)\,,
\end{equation}
where $y$ is
\begin{equation}\label{yy}
   y= u^\mu \partial_\mu \psi=\frac{ J^\mu\,\partial_{\mu}\psi}{b}\,.
\end{equation}

 By coupling (\ref{symmlagr}) to a gravity background, one can obtain the energy-momentum tensor by taking,
 as usual, the variation of $S$ with respect to a background (inverse) metric $g^{\mu\nu}$:
\begin{equation}\label{finaltmunu}
   T_{\mu\nu}=\left(y\,F_y-b\,F_b\right)u_\mu\,u_\nu +\eta_{\mu\nu}\left(F-b\,F_b\right)\,.
\end{equation}
On the other hand, from classical fluid dynamics, we also have
\begin{equation}\label{finaltmunucl}
   T_{\mu\nu}=\left(p+\rho\right)u_\mu\,u_\nu +\eta_{\mu\nu}p\,,
\end{equation}
from which we identify the pressure and density
\begin{equation}\label{ossia}
   \rho =y\,F_y-F\equiv y\,n-F\,\,\,,\,\,\,
p= F-b\,F_b\,.
\end{equation}
Comparing the two expressions of the energy-momentum tensor one can derive the relations between the
thermodynamical functions and the field-theoretical quantities  (see \cite{Dubovsky:2011sj} for a complete
review).
 In particular, it
turns out that the quantity $y$ defined in eq. (\ref{yy}) coincides with the chemical potential  $\mu$.
We conclude that the Lagrangian density of a perfect fluid is a function of $s$ and $\mu$
\begin{equation}\label{concl}
    F=F(s,\mu)\,.
\end{equation}

The  results presented here can be straightforwardly generalized to the supersymmetric case, where the comoving
coordinates $\phi^I$ and phase $\psi$ are extended to superfields. This was given in
\cite{Andrianopoli:2013dya}.

The Lagrangian (\ref{concl}) used above depends on the two Poincar\'e-invariant variables $b$ and $y$, but
 at first sight it
does not seem to allow for the presence of a kinetic term for the dynamical field $\psi$, namely $X=
\partial_\mu \psi
\partial^\mu \psi$. This term  respects the translational invariance (\ref{psitrans}),
 and was actually considered in \cite{Nicolis:2011cs},  but fails to satisfy the \emph{chemical-shift symmetry}
 (\ref{extsym}). However, as we are going to see, the kinetic term for $\psi$ is in fact included in $y^2$.

To show this, let us observe that the Poincar\'e invariants one can build from the fields $\partial_\mu \phi^I$
and $\partial_\mu \psi$ are given by $B^{IJ}$, $y$, $X$ together with the variables $Z^I \equiv \partial_\mu \psi
\partial^\mu \phi^I$.
Under chemical-shift symmetry they transform as
\begin{eqnarray}\label{SSN4A}
&& \delta B^{IJ}= 0\,,  \nonumber \\
&& \delta X = 2 \partial_I f Z^I\,, \nonumber \\
&& \delta y =  0 \,, \nonumber \\
&& \delta Z^I =  \partial_J f B^{IJ}\,.
\end{eqnarray}
 We note that $B^{IJ}$ and $y$ are invariant under the chemical-shift symmetry, while the
  other two quantities $X$ and $Z^I$ are not. However
   we can construct out of them  a new invariant $I( B,  X, Z)$:
\begin{equation}
I(B\,, X\,, Z)={X}-{Z}^I\,{Z}^J\,({B}^{-1})_{IJ}\,,\label{invar}
\end{equation}
which is inert under chemical shift symmetry as well.
In fact the new invariant (\ref{invar}) actually coincides with $- y^2$, since
\begin{eqnarray}
I( B\,, X\,, Z)&=& \partial_\mu \psi \partial_\nu \psi \left(\eta^{\mu\nu} -\partial^\mu \phi^{I}\partial^\nu\phi^{J}B^{-1}_{IJ}\right)\nonumber\\
&=&\partial_\mu \psi \partial_\nu \psi \left(-  u^\mu   u^\nu\right) = -y^2 \label{invary}
\end{eqnarray}
by virtue of the identity (see for example \cite{Dubovsky:2011sj,Nicolis:2011cs})
\begin{equation}
\partial_\mu \phi^{I}\partial_\nu\phi^{J}B^{-1}_{IJ}= \eta_{\mu\nu}+ u_\mu u_\nu\,.
\end{equation}
In terms of $X,Z,B$ a Poincar\'e invariant Lagrangian, enjoying the symmetries (\ref{symm1}) - (\ref{symm3}) and (\ref{psitrans}), and generalizing  (\ref{concl}) can be constructed.
It should  be an $SO(d)$-invariant functional of the
fields $B^{IJ},X,Z^I$:
\begin{equation}
 S= \int \mathcal{F}(B,X,Z)\,d^{d+1}x\,.\label{genact}
 \end{equation}
A Lagrangian of this kind is useful to describe physical situations where the chemical-shift symmetry is
spontaneously broken, and a particular instance of it was considered, for example, in \cite{Nicolis:2011cs} to describe a
superfluid at $T=0$.

The restricted form $F(b,y)$ is however required in all the cases where the chemical-shift invariance of the
Lagrangian is expected. In this case, the fields $X,Z^I$ can only appear  in the invariant combination
(\ref{invar}), and one recovers the standard form of the Lagrangian $F(b,y)$. We will discuss a possible
application of this extended formalism below, when we consider the case of the superfluid phase
transition, where the chemical-shift symmetry is spontaneously broken, together with its constant part (\ref{psitrans}).

Let us develop the dynamics deriving from the action principle (\ref{genact}). The variation of the
Poincar\'e-invariant fields under a generic variation of $\phi^I$ and of $\psi$ are
\begin{eqnarray}\label{SSN5}
&& \delta B^{IJ}  =- 2 d\phi ^I\wedge \star \, d \delta \phi^J \,,  \nonumber \\
&& \delta X = -2\, d\psi \wedge \star  \, d \delta \psi  \,, \nonumber \\
&& \delta Z ^I= -d \delta \psi\wedge \star \, d\phi^I   - d\psi \wedge \star  d \delta \phi ^I\,. \nonumber
\end{eqnarray}
The variation of $y$ can be given in terms of the
variation of $B^{IJ},X,Z^I$. The following equations of motion are obtained:
\begin{align}\label{SSN6}
&d\star  J_I^{(1)}= d\Big[ 2\,\mathcal{F}_{B^{IJ}}  \star\,  d\phi^J
 + \mathcal{F}_{Z^I } \star  \, d\psi \Big] =0\,, \\
& d\star  j^{(1)}= d \Big[ 2 \mathcal{F}_X \star \, d\psi
+ \mathcal{F}_{Z^I } \star \, d\phi^I \Big] =0\,,
\end{align}
where $ \mathcal{F}_{B^{IJ}} = \partial \mathcal{F} / \partial {  B^{IJ}}\,, \mathcal{F}_X = \partial
\mathcal{F} /
\partial {X}\,,
\mathcal{F}_{Z^I }  = \partial \mathcal{F} / \partial {Z}^I  $ and we have introduced the two currents:
\begin{align}
\label{deltaphi}J_I^{(1)}&= 2\,\mathcal{F}_{B^{IJ}}d\phi^J
 +\mathcal{F}_{Z^I } d\psi\,,\\
\label{deltaphi2}j^{(1)}&= 2 \mathcal{F}_X d\psi
+ \mathcal{F}_{Z^I }  d\phi^I\,.
\end{align}
It is straightforward to verify that the above  quantities are  the Noether currents associated with the
constant translational symmetries $\phi^I\rightarrow \phi^I+c^I$ and $\psi\rightarrow \psi+c$.

It is interesting to write down the energy momentum tensor for the generalized Lagrangian $\mathcal{F}[B,X,Z]$.
The new energy-momentum tensor $\tilde T_{\mu\nu}$ is now
\begin{equation}
\tilde T_{\mu\nu}= \eta_{\mu\nu} \mathcal{F} -2\left( \frac{\partial \mathcal{F}}{\partial   X} \Omega_\mu \Omega_\nu + \frac{\partial \mathcal{F}}{\partial   Z^I}
\Omega_\mu \Pi^I_\nu + \frac{\partial \mathcal{F}}{\partial   B^{IJ}}\Pi^I_\mu \Pi^J_\nu\right)\,.\label{text}
\end{equation}
To recover  a Lagrangian enjoying invariance under
chemical-shift symmetry
\begin{equation}
\mathcal{F}[  B,  X,  Z] = { F}[  s,  y]\,,
\end{equation}
one has to consider the case where  the fields $B^{IJ}$, $X$ and $Z^I$  only appear in the invariant combinations
(\ref{invary}) and $b^2=\det(B)$. Consequently, the corresponding energy-momentum tensor is retrieved from
(\ref{text}) by using the relations $y=y(B,X,Z)$ and $s=b=\sqrt{\det(B^{IJ})}$


As already emphasized, the generalized Lagrangian $\mathcal{F}$ can be useful for describing
fluid dynamics. One possible application can be found in the description of the 2-fluid model for superfluidity,
in the same spirit as the approach  in \cite{Nicolis:2011cs}. Let us recall few well known properties of
the Helium superfluid phase transition:
\begin{itemize}
\item Above the critical temperature $T_C$ the fluid has a normal behavior and is invariant under the
chemical-shift symmetry \cite{Dubovsky:2011sj}. It is described in terms of the comoving coordinates
$\phi^I(x)$ and by the $U(1)$-phase field $\psi(x)$.

 On the other hand, below the critical temperature the chemical-shift symmetry is spontaneously broken,
giving rise to the superfluid. In particular, at $T=0$, the superfluid is completely described in terms of
$\psi$.
\item One can consider, following \cite{Nicolis:2011cs}, an isotropic and homogeneous background where
\begin{eqnarray} \psi=y_0 t\,,\quad \phi^I = b_0^{1/3} x^I\,.\label{bkd}
\end{eqnarray}
It corresponds to taking a configuration where the fields $\phi^I$ are comoving with the normal fluid part
(which is at rest in this background), the superfluid field $\psi $ being in relative motion with respect to it.
 Note that the loss of interactions between the two fluids is expressed by the property that
$Z^I = \partial_\mu \psi \partial^\mu \phi^I =0$ in the background.

\item Small perturbations about the background    (\ref{bkd}):
\begin{eqnarray} \psi=y_0( t + \pi^0(x))\,,\quad \phi^I = b_0^{1/3}( x^I+ \pi^I(x))\label{overbkd}
\end{eqnarray}
introduce a small interaction term $Z^I \neq 0$.
Note that the quantity $B^{-1}_{IJ}Z^I Z^J = \epsilon$ stays small in this regime, even if $\phi^I \to 0$ as $T\to 0$.
\end{itemize}
Given these considerations, we can make use of the relation (\ref{invar}) to observe that at very low
temperatures the quantity
\begin{equation}
y^2= -X + B^{-1}_{IJ}Z^I Z^J = -X+ \epsilon
\end{equation}
approaches the value $X$, which is not invariant under the chemical-shift symmetry.   In this regime the
Lagrangian $F(b,y)$ can be expanded in powers of $\epsilon$ around the background, neglecting contributions
$\mathcal{O}(\epsilon^k)$, for a finite $k$. The effective Lagrangian at low temperatures, which is no longer
invariant  under chemical-shift symmetry, becomes then of the kind $\mathcal{F}[X,Z^I,B^{IJ}]$, and
reduces to $\mathcal{F}[X]$ for $T\to 0$.

Let us finally observe that above the critical temperature the theory is still described by the Poincar\'e
invariant fields $X,B^{IJ},Z^I$  including the kinetic terms of $\phi^I$ and $\psi$ which, however, only appear
in the chemical-shift invariant combinations $b$, $y$ since $Z^I$ is not small anymore, so that the form
$F(s,y)$ of the Lagrangian is recovered at $T> T_C$. In the quantum regime, as in \cite{Endlich:2010hf},
the expansion around the background and by using the new invariant might lead to a natural reorganization of the
perturbation theory.
\\
\\
In the rest of this paper we give the supersymmetric extension of this model, following the approach developed
in \cite{Andrianopoli:2013dya}.
As discussed there, the extension to
$(d+1)$-dimensional  superspace of the fields $d\phi^I$ and $d\psi$ is given by
\begin{equation}\label{SSN1}
\Pi^I = d \phi^I + \frac i 2 \bar \theta \Gamma^I  \, d\theta\,,
\quad\quad
\Omega = d \psi +i\bar \tau \, d\theta\,.
\end{equation}
Here $\Gamma^I$ are the Clifford algebra $\Gamma$-matrices in $d+1$-dimensions, with the index $I=1,\dots,d$
running on the spatial components only of $\Gamma^a$, $a=0,1,\cdots,d$, while $\theta $, $\tau$ and $d\theta$
denote the matrix form of the \emph{Majorana} spinors in the $m=2^{d/2}$-dimensional spinor representation of
${\rm SO}(d,1)$. \footnote{For  \emph{Majorana-Weyl} spinors, the dimension of the  representation is instead
$m=2^{(d-1)/2}$.} The  1-forms
 $\Pi^I$ and $\Omega$ are invariant under the following supersymmetry transformations of the fundamental
 fields $\phi^I,\theta^\alpha$:
\begin{equation}\label{SSN2}
\delta \phi ^I= - i \bar\epsilon \Gamma^I \theta\,, \quad \delta \theta = \epsilon\,.
\end{equation}
One can also  generalize the chemical-shift symmetry of the bosonic case by setting
\begin{equation}\label{gcss}
  \delta \psi = f(\phi,\theta),\quad\quad \delta \Omega = \Pi^I \,  \partial_{\phi^I} f\,,
\end{equation}
while $\Pi^I$ remains invariant. Equation (\ref{gcss}) also implies
$$
\delta \tau_\alpha = - D_\alpha f(\phi, \theta)\,,
$$
where $D_\alpha = \partial_{\theta^\alpha} -\frac i 2\bar \theta^\beta \Gamma^I_{\beta\alpha} \partial
_{\phi^I}$.

In terms of the above  variables, we can build the following quantities that generalize the corresponding
bosonic Poincar\'e-invariant fields:
\begin{eqnarray}\label{SSN4}
&& B^{IJ}=- \Pi^I \wedge  \star  \Pi ^J=\Pi^I_\mu\Pi ^{J\mu}\, d^{d+1} x=\hat{B}^{IJ}\,d^{d+1} x\,,
\nonumber\\
&& X = -\Omega \wedge \star  \Omega = \Omega_\mu g^{\mu\nu} \Omega_\nu d^{d+1}x=\hat{X}\,d^{d+1}x\,, \nonumber \\
&& Y = \Omega \wedge \Pi ^1 \cdots \wedge \Pi^d=
\frac 1{d!}\epsilon^{\mu\nu_1\cdots\nu_d}\epsilon^{I_1\cdots I_d}
\Omega_\mu \Pi^{I_1}_{\nu_1}\cdots \Pi^{I_d}_{\nu_d}\ d^{d+1}x=\hat{Y}\,d^{d+1}x\,, \nonumber\\
&& Z^I = -\Omega \wedge \star  \Pi ^I= \Omega_\mu g^{\mu\nu} \Pi^I_\nu  d^{d+1}x=\hat{Z}^I\,d^{d+1}x\,.
\end{eqnarray}
We denote by the same letter, though with a hat on the top, the corresponding factor multiplying  the volume
form $d^{d+1}x$, for instance $\hat B^{IJ} =  \Pi^I_\mu g^{\mu\nu} \Pi^J_\nu$. Note in particular that
\cite{Dubovsky:2011sj}:
\begin{equation}
\hat b  =
\sqrt{\det \hat B^{IJ}}\label{b}\,.
\end{equation}
 They are invariant under supersymmetry transformations and transform as
\begin{eqnarray}\label{SSN4A}
&& \delta B^{IJ}= 0\,,  \nonumber \\
&& \delta X = -2\, \Omega \wedge \star  \delta \Omega = -2 \, \partial_I f\, \Omega\wedge\star  \Pi^I= 2 \partial_I f Z^I\,, \nonumber \\
&& \delta Y =  0 \,, \nonumber \\
&& \delta Z^I = -\delta \Omega \wedge \star  \Pi ^I= -\partial_J f\, \Pi^J \wedge \star  \Pi ^I = \partial_J f B^{IJ}\,,
\end{eqnarray}
under the chemical-shift symmetry. We note that $B^{IJ}$ and $Y$ are invariant under the chemical shift
symmetry, while the other two quantities $X$ and $Z^I$ are not. The supersymmetric generalization of
the invariant (\ref{invary}) takes now the form
\begin{equation}
I(\hat B\,,\hat X\,,\hat Z)=\hat{X}-\hat{Z}^I\,\hat{Z}^J\,(\hat{B}^{-1})_{IJ}\,.\label{invar2}
\end{equation}
The new invariant (\ref{invar}) actually coincides with $-\hat Y^2/{\hat b^2}$. This was illustrated  at the bosonic level in eq. (\ref{invary}). To show that the bosonic result  extends
straightforwardly to the supersymmetric case we need the relation
\begin{equation}
\Pi^I_\mu \hat u^\mu  =0\,,
\end{equation}
where $\hat u^\mu$ is defined in terms of the 0-picture\footnote{In \cite{Catenacci:2010cs} the characterisation of
integral forms and theory quantum numbers are discussed.} part $J^{d|0}$ of the super-entropy current $J^{d|m}$
introduced in \cite{Andrianopoli:2013dya}:
\begin{equation}
\hat u^\mu = \hat b^{-1} (\star J^{d|0})^\mu\,.
\end{equation}

The Lagrangian describing the dynamics of the supersymmetric fluid should then be an $SO(d)$-invariant
functional of the fields $B^{IJ},X,Z^I$:
\begin{equation}
 S= \int \mathcal{F}(\hat B,\hat X,\hat Z)\,d^{d+1}x\,,
 \end{equation}
 which is an obvious generalization of the
bosonic case.

The variations of the superfields $B^{IJ},X,Z^I$ under a generic variation of the bosonic fields $\phi^I$ and
$\psi$ are
\begin{eqnarray}\label{sSSN5}
&& \delta B^{IJ}  =- 2 \Pi ^I\wedge \star  d \delta \phi^J \,,  \nonumber \\
&& \delta X = -2\, \Omega \wedge \star  d \delta \psi  \,, \nonumber \\
&& \delta Z ^I= -d \delta \psi\wedge\star  \Pi^I   - \Omega \wedge \star  d \delta \phi ^I\,,
\end{eqnarray}
and the following equations of motion are obtained:
\begin{align}\label{sSSN6}
&d\star  J_I^{(1)}= d\Big[ 2\,\mathcal{F}_{B^{IJ}}  \star  \Pi^J +
 \mathcal{F}_{Z^I } \star  \Omega \Big] =0\,, \\
& d\star  j^{(1)}= d \Big[ 2 \mathcal{F}_X \star  \Omega
+ \mathcal{F}_{Z^I } \star  \Pi^I \Big] =0\,,
\end{align}
where $ \mathcal{F}_{B^{IJ}} = \partial \mathcal{F} / \partial {\hat B^{IJ}}\,, \mathcal{F}_X = \partial
\mathcal{F} /
\partial \hat{X}\,,
\mathcal{F}_{Z^I }  = \partial \mathcal{F} / \partial \hat{Z}^I  $ and we have introduced the two currents:
\begin{align}
\label{sdeltaphi}J_I^{(1)}&= 2\,\mathcal{F}_{B^{IJ}}\Pi^J +
 \mathcal{F}_{Z^I } \Omega\,,\\
\label{sdeltaphi2}j^{(1)}&= 2 \mathcal{F}_X \Omega
+ \mathcal{F}_{Z^I }  \Pi^I\,.
\end{align}
Furthermore, a general variation  of the fermionic fields $\theta$ and $\tau$, gives
\begin{align}\label{sSSN7}
& \delta B^{IJ} = -i \Pi^I \wedge \star  (\delta\bar \theta \Gamma^J d\theta + \bar\theta\Gamma^J d \delta \theta)  \,,  \nonumber \\
& \delta X = 2\,i\Omega \wedge \star  (\delta \bar\tau d\theta +\bar \tau d \delta \theta)  \,, \nonumber \\
& \delta Z ^I= i (\delta \bar\tau d\theta + \bar\tau d \delta \theta) \wedge \star  \Pi^I  +
\frac i 2 \Omega \wedge \star (\delta\bar \theta \Gamma^I d\theta + \bar\theta\Gamma^I d \delta \theta) \,,
\end{align}
so that the corresponding \emph{fermionic} equations of motion are
\begin{align}
J_I^{\mu}\,\Gamma^I\partial_\mu\theta+\eta_C\,j^\mu\,\partial_\mu \tau=0\,,
\end{align}
where $\eta_C$ is the sign appearing in the relation $\bar{\tau}d\theta=\eta_C\,\bar{d\theta}\tau$ and depends
on the property of the charge conjugation matrix in $(d+1)$-dimensions.

The link between the above relations and the field equations given in \cite{Andrianopoli:2013dya} is easily retrieved by
recalling that $\hat Y= \hat Y(\hat X,\hat Z,\hat B)$ through the relation (\ref{invar2}) and $\hat b=
\sqrt{\det(\hat B^{IJ})}$.

For completeness, we also give the general expression of the supercurrent:
\begin{align}
j_S =J_I \Gamma^I\, \theta + \eta_C j \,\tau\,.
\end{align}

The present formalism can provide a framework for studying supergravity/ supersymmetric fluid correspondence
along the lines of \cite{Bhattacharyya:2008jc}. In particular, it might provide the natural ground for studying
the dissipative effects as in \cite{Kontoudi:2012mu,Hoyos:2012dh,Erdmenger:2013thg} and a systematically
computation of entropy current corrections \cite{Bhattacharyya:2008xc,Loganayagam:2008is,Plewa:2013rga}.

\section*{Acknowledgements}
This work was partially supported by the Italian MIUR-PRIN contract 2009KHZKRX-007 ``Symmetries of the Universe
and of the Fundamental Interactions''. We would like to thank V. Penna and G. Policastro for useful discussions.

\end{document}